\documentclass[superscriptaddress,preprintnumbers,amsmath,amssymb,nofootinbib]{revtex4}

\usepackage{graphicx}
\usepackage{dcolumn}
\usepackage{bm}

\begin{document}

\preprint{WU-AP/275/07}

\title{Astrophysical Implications of Equation of State for Hadron-Quark
Mixed Phase: \\ 
Compact Stars and Stellar Collapses}

\author{Ken'ichiro Nakazato}
 \email{nakazato@heap.phys.waseda.ac.jp}
 \affiliation{%
Department of Physics, Waseda University,\\
3-4-1 Okubo, Shinjuku, Tokyo 169-8555, Japan
}%

\author{Kohsuke Sumiyoshi}
\affiliation{Numazu College of Technology, Ooka 3600, Numazu, Shizuoka 410-8501, Japan
}%

\author{Shoichi Yamada}%
 \altaffiliation[Also at ]{Advanced Research Institute for Science \& Engineering, Waseda University, 3-4-1 Okubo, Shinjuku, Tokyo 169-8555, Japan.}
 \affiliation{%
Department of Physics, Waseda University,\\
3-4-1 Okubo, Shinjuku, Tokyo 169-8555, Japan
}%

\date{\today}

\begin{abstract}
 We construct an equation of state including the hadron-quark phase
 transition. The mixed phase is obtained by the Gibbs conditions for
 finite temperature. We adopt the equation of state based on the
 relativistic mean field theory for the hadronic phase taking into
 account pions. As for the quark phase, the MIT bag model of the
 deconfined 3-flavor strange quark matter is used. As a result, our
 equation of state is thermodynamically stable and exhibits
 qualitatively the desired properties of hadron-quark mixed matter, such
 as the temperature dependence of the transition density. The pions
 raise the transition density because they make the equation of state
 softer. Using the equation of state constructed here, we study its
 astrophysical implications. The maximum mass of compact stars is
 investigated, and our equation of state is consistent with recent
 observations. We also compute the collapse of a massive star with 100
 solar masses ($M_\odot$) using our equation of state and find that the
 interval time from the bounce to the black hole formation becomes
 shorter for the model with pions and quarks. The pions and quarks
 affect the total energy of the emitted neutrinos because the duration
 time of the neutrino emission becomes shorter. The neutrino luminosity
 rises under the effect of pions since the density of the proto-neutron
 star becomes high.
\end{abstract}


\maketitle

\section{Introduction} \label{intro}
It has been theoretically suggested that hadronic matter undergoes a
deconfinement transition to quark matter at high temperature and/or high
density. It is also inferred that the transition occurs inside compact
stars \cite{itoh70, witten84, weber05}. Stars with a quark matter central
region and a hadronic matter mantle are called hybrid stars. While the
equation of state (EOS) for the hadron-quark mixed phase is not yet fully
understood, structures and maximum masses of the hybrid stars have been
well studied \cite{glende92, drago99, burgio02, mariero04, blaschke05,
nicotra06, baldo07, sandin07}, for review
Ref.~\onlinecite{glende02}. For the phase transition of a single
substance, like a liquid-vapor transition of H$_2$O, it is simple to
treat the mixed phase. On the other hand, in the hadron-quark
transition, the substance is composed not only of $u$ quarks but also of
$d$ quarks \cite{glende92}. Incidentally, some authors use the simple
Maxwell construction for the studies of hybrid stars \cite{mariero04,
nicotra06}. In this scheme, Gibbs conditions for two components are not
satisfied completely and EOS's of two phases are connected from the plot
of pressure versus chemical potential.

This transition may play an important role in the gravitational collapse
of a star, such as a core collapse supernova. In addition, there is a
possibility of forming a black hole by the transition because the
compact stars have maximum masses. In this case, the EOS with finite
temperature and including neutrinos is needed whereas the EOS can be
calculated under zero temperature and neutrino-less $\beta$ equilibrium
for hybrid stars. The inclusion of neutrinos is needed because they will
be trapped at a very high density as the hadron-quark transition
occurs. Moreover, for the gravitational collapse of a star, the EOS
should be constructed in a consistent method in the wide range of
density. Recently, dynamical simulations of the stellar core collapse
including the transition have been done; however, their EOS is obtained
at zero temperature and they adopt the Maxwell construction for mixed
phase \cite{gentile93, yasu07}. So far, the stellar collapse including
the hadron-quark phase transition with a finite temperature EOS and
neutrinos has not been studied.

A study on the mixed phase and hot hybrid stars using Gibbs conditions
for finite temperature with neutrinos exists; however, it does not
compute the dynamics of the stellar collapse and its EOS does not
include low density regime where heavy nuclei appear \cite{drago99}. In
Ref.~\onlinecite{drago99}, the relativistic mean field model of the
Walecka type is adopted for the EOS of hadronic matter. This model does
not take into account the non-linear self-coupling terms of $\sigma$
meson and $\omega$ meson, which are essential to reproduce the
properties of nuclei quantitatively and of dense matter in a reasonable
manner \cite{suga94}. In addition, pions are not included in its
hadronic phase while they may affect the transition density. The color
dielectric model is used to describe quark matter. We will revisit this
model in Sec.~\ref{eos} comparing with our model.

In this study, we construct the EOS of hadron-quark mixed matter
including neutrinos for finite temperature and perform the computations
of the stellar collapse and black hole formation. An EOS by
\citeauthor{shen98a} (Shen EOS) \cite{shen98a, shen98b} and the MIT bag
model \cite{bag74} are adopted for nuclear matter and quark matter,
respectively. Shen EOS is based on the relativistic mean field theory
which includes the non-linear self-coupling terms and takes into account
non-uniform matter. It is noted that the table of EOS covers a wide
density and temperature range and has been often used in astrophysical
simulations. Contributions of thermal pions are added to the original
Shen EOS in the current study to examine its effect in the minimum model
which assumes that an effective mass of pions is equal to their rest
mass in vacuum. Incidentally, hyperons are not included in our hadronic
EOS, however, we are planning to investigate their effects in future
work \cite{self08b}. The MIT bag model is characterized by a parameter
$B$, which is called the bag constant and related to the degree of
interaction. We calculate the EOS for several values for $B$ and
investigate its dependence. We show that our EOS is thermodynamically
stable because Gibbs conditions for each particle species are completely
satisfied for the mixed phase. After examining the neutron star
properties, we apply our EOS to the numerical simulation of the stellar
core collapse. It is noted that neutrinos are treated consistently in
our simulation. Thus we can evaluate the total energy of the neutrinos
emitted from the collapse. We show that the hadron-quark transition
affects the dynamics and neutrino emission of the collapse.

In this paper, we perform the computations only for one progenitor model
and limited cases of EOS's are used for the hadronic phase and quark
phase. This paper is the first stage of the project and we are preparing
to study the progenitor dependence \cite{self08a} and EOS dependence
\cite{self08b} of the collapse and neutrino emission. The final goal of
our study is to investigate the effects of the phase transition on the
stellar collapse and the possibilities to probe the EOS of hot dense
matter from the emitted neutrinos, as demonstrated for pure hadronic
matter in Refs.~\onlinecite{sumi06, sumi07}.

The organization of this paper is as follows. We devote Sec.~\ref{form}
to the formulation of the EOS for the hadron-quark mixed phase. We also
introduce EOS's of the pure hadronic and quark matter and present 
treatments of muons. In Sec.~\ref{result}, we assess our EOS and discuss
the  dependence on the bag constant, $B$. As astrophysical implications,
we  evaluate the maximum mass of the compact stars and investigate the
effects of hadron-quark phase transition on the stellar core
collapse. Finally, a summary is given in Sec.~\ref{summary}.

\section{Formulations} \label{form}
In this section, we construct the EOS for hadron, quark and their
mixture as a function of baryon mass density $\rho_B$, electron fraction
$Y_e$ and temperature $T$. The EOS calculated in this study covers
10$^{5.1}$~g~cm$^{-3}$~$\leq\rho_B\leq$~10$^{17}$~g~cm$^{-3}$,
$0\leq Y_e \leq0.56$ and 0~MeV~$\leq T \leq$~400~MeV. The lower limit of
$\rho_B$ and the range of $Y_e$ are the same as those of the Shen EOS,
which is adopted for the pure hadronic matter in our model. We assume
that the pure hadronic matter exists under the transition density and
the pure quark matter exists for a higher density than that of the end
point of the mixed phase. In the following, after we introduce the EOS
for pure hadronic matter and pure quark matter, we give the formulations
of the EOS for mixed matter. Finally, we also describe the treatment of
muons.

\subsection{EOS for pure hadronic matter} \label{hadron}
We adopt the Shen EOS \cite{shen98a, shen98b} for the pure hadronic
matter. This EOS is based on the relativistic mean field theory and the
table of EOS covers a wide baryon-mass density range,
$\rho_B=10^{5.1}$~-~10$^{15.4}$~g~cm$^{-3}$, and temperature range,
$T=0$~-~100~MeV. An inhomogeneity of the matter is also taken into
account using the Thomas-Fermi approximation because heavy nuclei appear
in $T\lesssim15$~MeV and densities below the nuclear matter
density. It is noted that the density where the hadron-quark transition
occurs is higher than the nuclear matter density for $T\lesssim15$~MeV
in our model. Thus the hadronic matter at the transition point is
composed of dissociated protons and neutrons. It is also noted that we
extend the EOS to $T>100$~MeV by a method fully consistent with the Shen
EOS performing the calculations based on the relativistic mean field
theory. In this temperature regime, we can regard that the nuclear
matter is uniform and the scheme for computations is given in
Refs.~\onlinecite{sumi94, sumi95a} in detail. The high temperature
regime will be revisited in Sec.~\ref{eos}.

We add the effects of charged pions, $\pi^\pm$, and neutral pions,
$\pi^0$, to the EOS for the pure hadronic matter. The original Shen EOS
is based on the mean field theory without the pions. It is noted that
the pion field vanishes in the mean field approximation because it is a
pseudoscalar particle \cite{glende83a}. In the realistic dense matter,
pions may exist though large uncertainties still exist. Thus we examine
their qualitative effects by treating them in the minimum model. Here we
employ the method in Refs.~\onlinecite{glende83a, glende83b, glende85}
under the assumption that an effective mass of pions is equal to their
rest mass in vacuum. Strictly speaking, pions feel an repulsive
potential in the nucleons and their effective mass gets larger than that
in vacuum. In this case, the pion population is suppressed. Thus our
assumption corresponds to one extreme case and the situation without
pions does the other extreme case \cite{glende83a, glende83b}. The
realistic condition should be between these two cases.

The pions are handled as an ideal boson gas and the chemical potentials
of the neutral and charged pions are determined as follows. The chemical
potential of the neutral pion is always $\mu_{\pi^0}=0$ because it is a
self-conjugate particle and created by the pair process of two
photons. As for the chemical potentials of the charged pions, we set
them as $\mu_{\pi^-}=\mu_n-\mu_p$ and $\mu_{\pi^+}=-\mu_{\pi^-}$, where
$\mu_p$ and $\mu_n$ are the chemical potentials of proton and neutron, 
respectively. When the charge chemical potential, $\mu_n-\mu_p$, exceeds
the rest mass of charged pions, $m_{\pi^\pm}$, at high densities, the
charged pions are condensed and the charge chemical potential decreases
to their rest mass. In fact, there is such a regime in the original Shen
EOS table. It is well known that this threshold is sensitive to the
density dependence of the symmetry energy. We remark that the density
dependence is strong in the relativistic nuclear many body frameworks
such as the relativistic mean field theory for Shen EOS, as compared
with the non-relativistic counter part \cite{sumi95b}. 

In the following, we replace the original Shen EOS given as a function
of baryon mass density $\rho_B$, proton fraction $Y_p$ and temperature
$T$ with our hadronic EOS given as a function of $\rho_B$, $Y_C$ and
$T$. $Y_C$ is a charge fraction and defined as
\begin{equation}
Y_C=\frac{n_p - n_{\pi^-}}{n_B},
\label{yc}
\end{equation}
where $n_B$ and $n_p$ are the baryon number density and the number
density of protons, respectively. $n_{\pi^-}$ is a net number density of
the charged pions, which is the difference of the number density of
$\pi^-$ to that of $\pi^+$, and calculated from the Bose-Einstein
distribution function:
\begin{equation}
n_{\pi^-}=\displaystyle \frac{1}{h^3} \int^\infty_0 \left[ \frac{1}{\left\{\exp \left(\sqrt{m_{\pi^\pm}^2 c^4 + p^2 c^2} - \mu_{\pi^-} \right) / (k_\mathrm{B}T) \right\} -1} - \frac{1}{\left\{\exp \left(\sqrt{m_{\pi^\pm}^2 c^4 + p^2 c^2} + \mu_{\pi^-} \right) / (k_\mathrm{B}T)\right\}-1} \right] 4\pi p^2\mathrm{d}p,
\label{pith}
\end{equation}
where $h$, $k_\mathrm{B}$ and $c$ are the Planck constant, the Boltzmann
constant and the velocity of light, respectively. When the pions are not
condensed, we find $\mu_{\pi^-}$ and the corresponding state of the
nucleons in the Shen EOS for given $\rho_B$, $Y_C$ and $T$. In this
process, we determine the relations of $n_p$, $\mu_p$ and $\mu_n$ from
the Shen EOS. Using $\mu_{\pi^-}$ found above, the pressure and energy
density of the pions are calculated.

On the other hand, in case of the pion condensation, we fix the chemical
potential as $\mu_{\pi^-}=m_{\pi^\pm} c^2$ and calculate the number
density of the ``thermal'' pions, $n_{\pi^-}^\mathrm{th}$, in
Eq. (\ref{pith}). The pressure and energy density of ``thermal'' pions
are given from the Bose-Einstein distribution function as in the case
without the pion condensation. At the same time, we can determine the
EOS of nucleons for given $\rho_B$ and $T$ using the Shen EOS under the
condition of $\mu_n-\mu_p=m_{\pi^\pm} c^2(=\mu_{\pi^-})$. Having $Y_p$
and $n_p$ fixed, we can get the net number density of the pions,
$n_{\pi^-}$, for given $Y_C$ using Eq. (\ref{yc}). Here we set the
number density of the ``condensed'' pions as
$n_{\pi^-}^\mathrm{cond}=n_{\pi^-}-n_{\pi^-}^\mathrm{th}$. These
condensed pions contribute not to the pressure but to the energy density
by their rest masses. We can determine the EOS including condensed pions
approximately.

Incidentally, when the electron-type neutrinos are trapped and in
equilibrium with other particles, their chemical potential should be
given as Eq.~(\ref{chemipa}), which is expressed later. For high density
regime where the hadron-quark transition occurs, neutrinos are fully
trapped and we show the results including trapped neutrinos in
Sec.~\ref{eos}. On the other hand, since neutrinos are not trapped at
least for the onset of gravitational collapse, we follow the time
evolution of neutrino distributions through neutrino reactions for
numerical simulations in Sec.~\ref{collapse}. In addition to the quark
degree of freedom, hyperons and kaons also may be important in the high
density regime. In particular, since hyperons will appear before the
hadron-quark phase transition \cite{glende85, balberg99}, we are
investigating their effects in future work \cite{self08b}.

\subsection{EOS for pure quark matter} \label{quark}
We adopt the MIT bag model \cite{bag74} as the pure quark matter. This
is a phenomenological model which describes the nature of the
confinement and the asymptotic freedom of quarks. In this model, free
quarks are confined in the ``bag'' and this ``bag'' has a positive
potential energy per unit volume \cite{bag74}. Thus the thermodynamical
potential is expressed as $\Omega_Q=\Omega_0+BV$,\footnote{Some authors
use a formula, which includes the lowest-order gluon interaction with
the coupling constant $\alpha_s$ \cite{gentile93, burgio02,
yasu07}. However, the validity is not guaranteed for the densities in
which we are interested because the perturbative approach is valid only
in the high energy limit. Hence, we do not take into account these
effects in this study.} where $V$ and $\Omega_0$ are the volume of the
``bag'' and the thermodynamical potential of free quarks as ideal
fermions, respectively. The bag constant, $B$, is a parameter
characterizing this model, and we investigate its dependence later. From
thermodynamical relations, we can calculate the number density $n_Q$,
pressure $P_Q$ and energy density $\varepsilon_Q$ for the quark matter
with the temperature $T$ as 
\begin{subequations}
\label{bag}
\begin{eqnarray}
n_Q & = & \sum_f \frac{g}{h^3} \int^\infty_0 \left(F^{+}_f(p) - F^{-}_f(p) \right)4\pi p^2\mathrm{d}p, \\
\label{bagn}
P_Q & = & \sum_f \frac{g}{h^3} \int^\infty_0 \frac{p^2 c^2}{3 \sqrt{m_f^2 c^4 + p^2 c^2}}\left(F^{+}_f(p) + F^{-}_f(p) \right)4\pi p^2\mathrm{d}p-B, \\
\label{bagp}
\varepsilon_Q & = & \sum_f \frac{g}{h^3} \int^\infty_0 \sqrt{m_f^2 c^4 + p^2 c^2} \left(F^{+}_f(p) + F^{-}_f(p)\right)4\pi p^2 \mathrm{d}p+B,
\label{bage}
\end{eqnarray}
\end{subequations}
where $F^{+}_f(p)$ and $F^{-}_f(p)$ represent the Fermi-Dirac
distribution functions for particle and antiparticle, respectively, and
they are expressed as
\begin{equation}
 F^{\pm}_f(p)=\frac{1}{\left\{\exp \left(\sqrt{m_f^2 c^4 + p^2 c^2} \mp \mu_f \right) / (k_\mathrm{B}T)\right\}+1}.
\label{fermi}
\end{equation}
The subscript $f$ denotes the flavor of quarks, and we take into account
three flavors, namely, $u$-, $d$- and $s$ quarks. The statistical weight
is $g=2\times3$. $m_f$ and $\mu_f$ are the mass and the chemical
potential of $f$ quark, respectively, and we adopt $m_u c^2=2.5$~MeV,
$m_d c^2=5$~MeV and $m_s c^2=100$~MeV in this study \cite{pdg06}.

As mentioned already, the EOS for the pure hadronic matter is given as
the function of the baryon mass density $\rho_B$, the charge fraction
$Y_C$ and the temperature $T$. We can rewrite the EOS of the MIT bag
model as the function of these three independent variables. For
convenience, we define the ``baryon'' number density and the ``baryon''
mass density of the quark matter as
\begin{subequations}
\label{bq}
\begin{eqnarray}
n_{B,Q} & = & \frac{n_Q}{3}, \\
\label{nbq}
\rho_{B,Q} & = & m_\mathrm{unit}\frac{n_Q}{3},
\label{rbq}
\end{eqnarray}
\end{subequations}
where the value of the atomic mass unit $m_\mathrm{unit}=931.49432$~MeV
is used as in the Shen EOS. Since the electric charges of $u$-, $d$- and
$s$ quarks are $\frac{2}{3}$, $-\frac{1}{3}$ and $-\frac{1}{3}$ of that
of a proton, respectively, the charge fraction of the quark matter is as
follows:
\begin{equation}
Y_{C,Q}=\frac{2n_u - n_d - n_s}{n_u + n_d + n_s}=\frac{2n_u - n_d - n_s}{n_Q},
\label{ypq}
\end{equation}
where $n_f$ is a number density of $f$ quarks. When the $\beta$
equilibrium is satisfied, $\mu_d$ is equal to $\mu_s$, which is
expressed later as Eq. (\ref{chemipd}). Thus we can determine $\mu_u$,
$\mu_d$ and $\mu_s$ for given $\rho_{B,Q}$, $Y_{C,Q}$ and $T$, and we
can express $P_Q$ and $\varepsilon_Q$ as functions of $\rho_{B,Q}$,
$Y_{C,Q}$ and $T$. We can also calculate other variables, such as the
entropy density $s_Q=\frac{P_Q + \varepsilon_Q}{T} + \frac{1}{T} \sum_f
n_f \mu_f$, the Helmholtz free energy per unit volume ${\cal
F}_Q=\varepsilon_Q-Ts_Q$ and the Gibbs free energy per unit volume
${\cal G}_Q={\cal F}_Q+P_Q$. We will use these variables later in this
paper.

\subsection{EOS for hadron-quark mixed phase} \label{mix}
Following Ref.~\onlinecite{glende92}, we first show the conditions of
the equilibrium between phases in the heat bath with the temperature
$T$. Here, we deal with hadronic matter, quark matter, electrons,
electron type neutrinos and mu-type leptons, whose treatment is stated
later in Sec.~\ref{muon}. We assume that the equilibrium is achieved not
only by the strong interactions but also by the weak interactions. This
is because only $u$- and $d$ quarks are deconfined from protons and
neutrons and $s$ quarks are created by the weak interactions. Moreover,
we assume that the neutrinos are completely trapped owing to high
density. In other words, the diffusion time scale is much longer than
the reaction time scale. Thus the following reactions are in chemical
equilibrium:
\begin{subequations}
\label{chemie}
\begin{eqnarray}
p + e^- & \longleftrightarrow & n + \nu_e, \\
n & \longleftrightarrow & u + 2d,  \\
p & \longleftrightarrow & 2u + d,  \\
d & \longleftrightarrow & u + e^- + \bar\nu_e, \\
s & \longleftrightarrow & u + e^- + \bar\nu_e, \\
u + d & \longleftrightarrow & u + s,
\end{eqnarray}
\end{subequations}
and the relations of the chemical potentials are given as
\begin{subequations}
\label{chemip}
\begin{eqnarray}
\label{chemipa}
\mu_p + \mu_e & = & \mu_n + \mu_\nu, \\
\label{chemipb}
\mu_n & = & \mu_u + 2\mu_d, \\
\label{chemipc}
\mu_p & = & 2\mu_u + \mu_d, \\
\mu_d & = & \mu_s.
\label{chemipd}
\end{eqnarray} 
\end{subequations}
It is noted that electrons reside in both phases and the chemical
potential for each phase coincides with each other. This is also true
for neutrinos. It goes without saying that two phases are also in
mechanical equilibrium. For simplicity, we ignore the surface tension
and the screening of the charged particles though they may affect the
EOS \cite{endo06}. Thus we require the condition,
\begin{equation}
P_H=P_Q,
\label{atsuryoku}
\end{equation}
where $P_H$ and $P_Q$ are the pressures of hadronic and quark phases,
respectively. It is noted that we do not need to take into account the
contribution of leptons in Eq. (\ref{atsuryoku}), because their chemical
potentials and pressures are the same for each phase.

Next, we relate the independent variables in each phase to the average
values using a volume fraction of the quark phase, $\chi$. For instance,
a pure hadronic phase corresponds to $\chi=0$ and a pure quark phase to
$\chi=1$. Then the baryon mass density of the mixed phase, $\rho_B$, is
written as
\begin{equation}
\rho_B=(1-\chi)\rho_{B,H}+\chi\rho_{B,Q},
\label{rhob}
\end{equation}
where $\rho_{B,H}$ is the baryon mass density of the hadronic phase and
$\rho_{B,Q}$ is that of the quark phase and given in Eq. (\ref{rbq}). We
note that, the charge fractions of both phases, $Y_{C,H}$ and $Y_{C,Q}$,
do not necessarily coincide with each other. Since the mixed phase is
charge neutral as a whole, we can relate these values with the electron
fraction of the mixed phase, $Y_e$, as
\begin{equation}
Y_e \rho_B=(1-\chi)Y_{C,H}\rho_{B,H}+\chi Y_{C,Q}\rho_{B,Q}.
\label{ye}
\end{equation}
It is noted that the charge neutrality is not required for each phase
independently \cite{glende92}. This fact gives an essential difference
from the phase transition of a single substance like a liquid-vapor
transition of H$_2$O. For instance, the pressure of the hadron-quark
mixed phase is not constant in an isothermal process. Now the
construction of the EOS for hadron-quark mixed phase is reduced to the
determination of the volume fraction, $\chi$, for given $\rho_B$,
$Y_e$ and $T$. Mathematically, it is equivalent to solving the system of
5 equations (\ref{chemipb}, \ref{chemipc}, \ref{atsuryoku}, \ref{rhob},
\ref{ye}) for 5 variables, namely, $\chi$, $\rho_{B,H}$, $Y_{C,H}$,
$\rho_{B,Q}$ and $Y_{C,Q}$. Here $\rho_B$ and $Y_e$ are given, $\mu_n$,
$\mu_p$ and $P_H$ are the functions of $\rho_{B,H}$ and $Y_{C,H}$, and
$\mu_u$, $\mu_d$ and $P_Q$ are the functions of $\rho_{B,Q}$ and
$Y_{C,Q}$.

\subsection{Treatment of muons} \label{muon}
As mentioned already, the treatment of muons and their anti-particles is
important because their compositions affect the EOS. In this study, we
examine two situations.

The first one corresponds to the EOS for stellar core collapse. Here, we
call it supernova (SN) matter. In the case of an ordinary supernova, it
is known that not only the net muon fraction but also the muon and
anti-muon fractions themselves are minor because the rest mass of the
muon is large ($m_\mu=105.66$~MeV). Then, they are omitted in
simulations for the ordinary supernova. On the other hand, we may not be
able to neglect their effects for EOS in case of black hole formation,
because the density and temperature would be higher than those for the
ordinary supernova \cite{sumi05, self07}.

During the core collapse and bounce, muons behave as follows. Since the
charged current reactions for mu-type leptons are not efficient before
neutrino trapping, mu-type neutrinos ($\nu_\mu$) and their
anti-particles ($\bar\nu_\mu$) are produced mainly by pair processes,
and their reactions are almost identical. Thus one can safely posit that
the net muon fraction is zero during the initial collapse before
neutrino trapping. When neutrinos are fully trapped, the charged current
reactions may become efficient. Therefore, muons and mu-type neutrinos
are in chemical equilibrium. $\nu_\mu$ and $\bar\nu_\mu$ cannot be
regarded as identical, and the net muon fraction can become non-zero.

In this study, in order to add the effects of muon and anti-muon pairs
in a tractable manner, we do not assume chemical equilibrium for muon
type leptons\footnote{We remark that the $\beta$ equilibrium assumed in
previous sections is for the electron-type leptons. This assumption is
not inconsistent with the treatment of muons.} but suppose that the net
muon fraction is zero (i.e., $\mu_\mu=0$). The approximation $\mu_\mu=0$
is reasonable because the pair population dominates in higher
temperature regime, where we cannot neglect the effects of muon-type
leptons for EOS in the black hole formation. Incidentally, this
assumption is consistent with the methods of our simulations stated in
Sec.~\ref{collapse}, which does not take into account the muon-charged
current reactions. In order to treat the muon-charged current reactions,
one has to solve the neutrino-transfer for 6 species, which requires
formidable computing resources. It is also necessary to prepare the EOS
table as a function of not only $Y_e$ but also $Y_\mu$, which makes the
setups further complicated. The muon-charged current reactions should be
included in future to assess their effects.

The second case of the treatment of muons is for so-called neutron star
(NS) matter. In this case, the neutrino less $\beta$ equilibrium at zero
temperature is achieved and the chemical potential of the muons,
$\mu_{\mu}$, is the same as that of electrons. Thus the relations of the
chemical potentials become as follows:
\begin{subequations}
\label{chemipns}
\begin{eqnarray}
\mu_e & = & \mu_n - \mu_p, \\
\mu_\mu & = & \mu_e.
\end{eqnarray} 
\end{subequations}
Moreover, the charge neutrality condition (\ref{ye}) is modified to 
\begin{equation}
(Y_e+Y_\mu) \rho_B=(1-\chi)Y_{C,H}\rho_{B,H}+\chi Y_{C,Q}\rho_{B,Q},
\label{yens}
\end{equation}
where $Y_\mu$ is a muon fraction.

\section{Results} \label{result}
In this section, we show the features of the EOS described in the
preceding section. We choose a mixed EOS with the hadronic matter
including pion and the quark matter of the bag constant
$B=250$~MeV~fm$^{-3}$ ($B^{1/4}=209$~MeV), as a reference model. This
value is adopted in another recent study on the hadron-quark phase
transition \cite{wang07}. Here we discuss the properties of EOS with the
applications to the maximum mass of compact stars and a stellar
collapse. We mainly show the results of the reference model for the NS
matter and the SN matter. Incidentally, results of the model without
pions are also shown, and the dependence on the bag constant is studied
for several topics. Further studies on stellar collapse will be reported
elsewhere \cite{self08a, self08b}.

\subsection{Assessments of EOS} \label{eos}
\begin{figure}
\begin{center}
\includegraphics{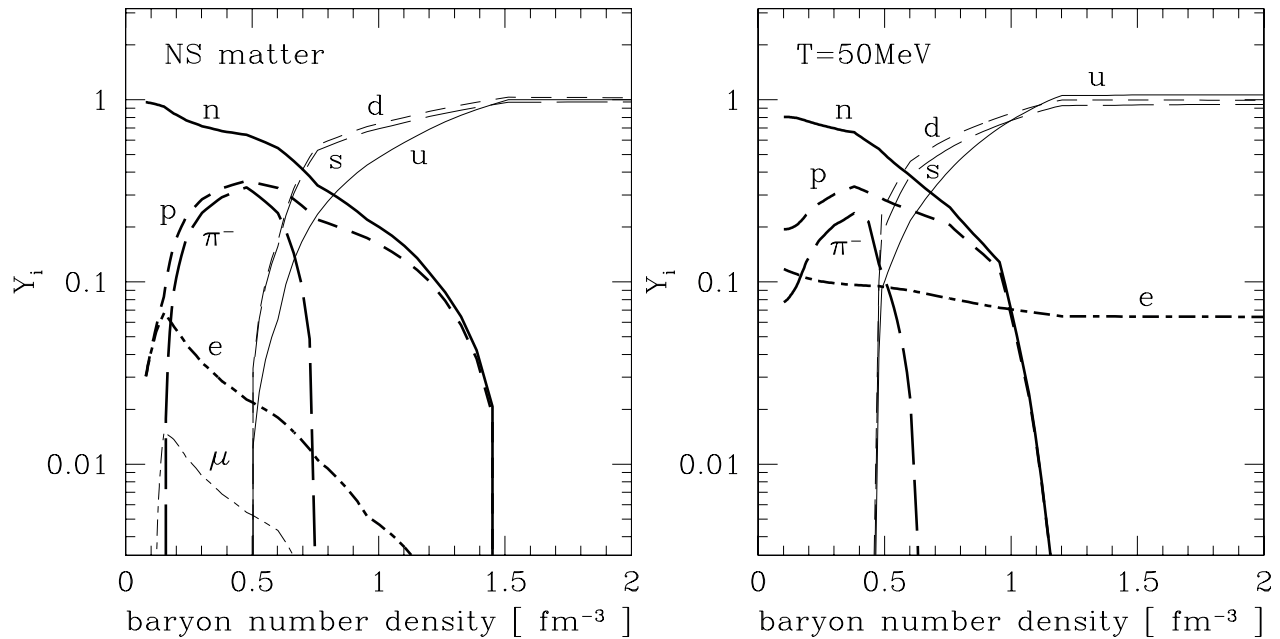}
\caption{Particle fractions for the reference model of NS matter
 (left) and SN matter with $T=50$~MeV and $Y_l=0.1$ (right). The
 particle fraction, $Y_i$, is defined as $\frac{n_i}{n_B}$, where $n_i$
 represents the number density of the particle $i$.}
\label{comp}
\end{center}
\end{figure}

First of all, we examine the composition of our EOS for NS matter
and SN matter to compare with other EOS's in the previous studies using
similar schemes \cite{glende85, glende92, drago99, nicotra06}. For SN
matter, we show the cases of $T=$~50~MeV with the electron type lepton
fraction $Y_l=0.1$, where $Y_l$ is defined as the sum of the electron
fraction, $Y_e$, and the electron-type neutrino fraction,
$Y_{\nu_e}$. Note that when the neutrinos are fully trapped, $Y_l$ is
conserved for each fluid element in the stellar core. In
FIG.~\ref{comp}, we show the particle fractions, $Y_i \equiv
\frac{n_i}{n_B}$, where $n_i$ represents the number density of the
particle $i$, for the reference models. When pions begin to condensate
in the NS matter, the proton fraction increases and the fractions of
electron and muon decrease \cite{glende85}. At the onset of the
hadron-quark phase transition, $d$ quarks have larger population than
those of $u$- and $s$ quarks not only for the NS matter \cite{glende92}
but also for the SN matter \cite{drago99}. In the pure quark phase, the
number ratio of $u$-, $d$- and $s$ quark is nearly equi-partition and
the relation, $Y_s < Y_u < Y_d$ holds for the NS matter. For the SN
matter with neutrino trapping, $u$ quarks have larger population than
those of $d$- and $s$ quarks \cite{drago99, nicotra06}. All of these
features are consistent with the previous results.

\begin{figure}[t]
\begin{center}
\includegraphics{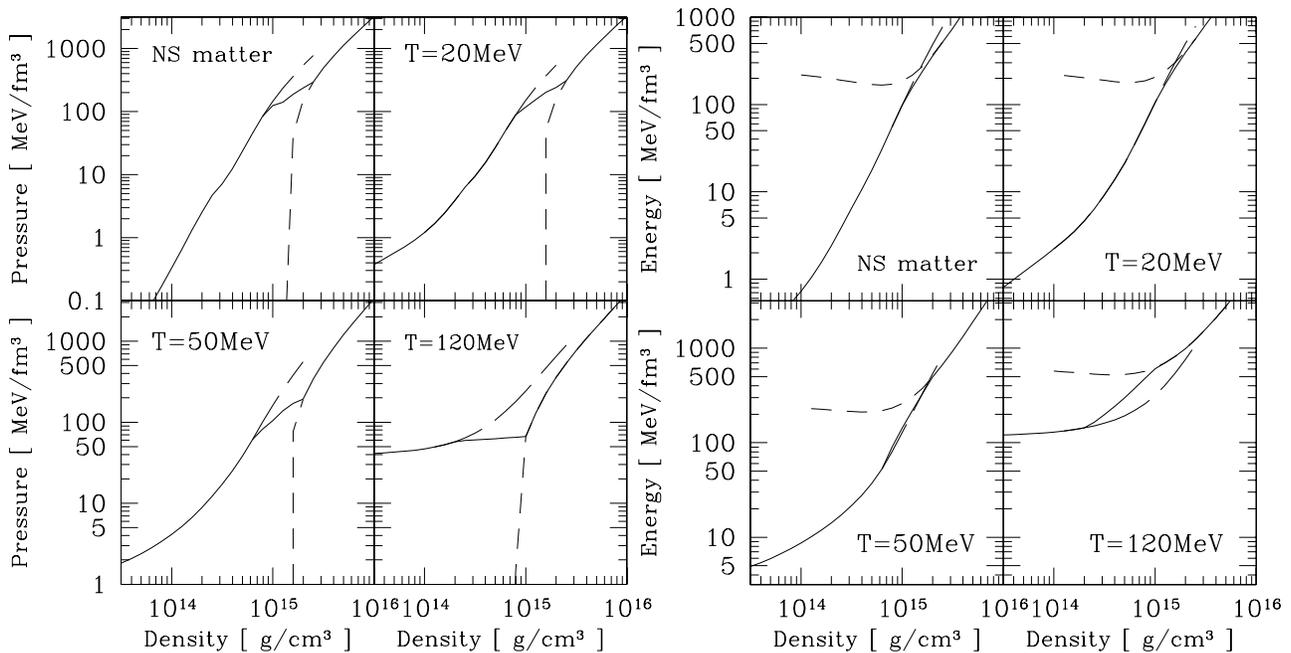}
\caption{EOS's for mixed matter (solid lines), pure hadronic matter
 (long-dashed lines) and pure quark matter (short-dashed lines). In all
 panels, the bag constant is chosen as $B=250$~MeV~fm$^{-3}$ for the
 quark model. The left and right panels show the pressure and the
 internal energy, respectively, as a function of the density. In both
 panels, the upper left, the upper right, the lower left and the lower
 right plots correspond to the reference models of NS matter, SN matter
 with $T=20$, 50 and 120~MeV, respectively. The electron type lepton
 fraction, $Y_l$, is fixed to 0.1 for all models of SN matter.}
\label{preene}
\end{center}
\end{figure}

\begin{figure}[t]
\begin{center}
\includegraphics{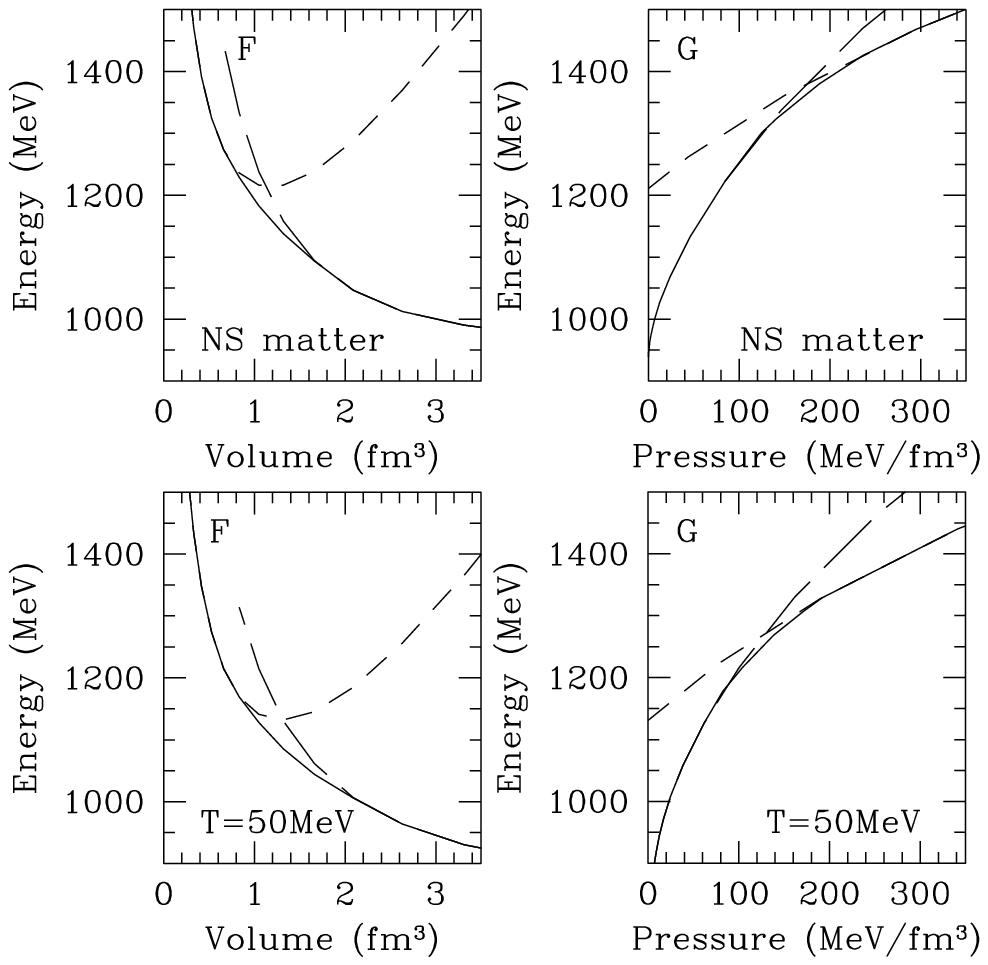}
\caption{EOS's for mixed matter (solid lines), pure hadronic matter
 (long-dashed lines) and pure quark matter (short-dashed lines). In all
 panels, the bag constant is chosen as $B=250$~MeV~fm$^{-3}$ for the
 quark model. The upper left and upper right panels show the Helmholtz
 free energies per baryon (F) as a function of a specific volume of the
 system and the Gibbs free energies per baryon (G) as a function of the
 pressure, respectively, for the reference model of NS matter. The lower
 two panels are the same as upper two panels but for the reference model
 of SN matter with $T=50$~MeV and $Y_l=0.1$.}
\label{fg}
\end{center}
\end{figure}

In FIG.~\ref{preene}, we compare the EOS for the reference model with
those for pure hadronic and quark matters, where the bag constant
$B=250$~MeV~fm$^{-3}$. This figure illustrates both for NS matter and SN
matter with $T=20$, 50 and 120~MeV and $Y_l=0.1$. As mentioned already,
the pressure of the hadron-quark mixed phase is not constant in an
isothermal process \cite{glende92}, which is well presented in our
models. In addition, this trend is shown also in the other previous work
\cite{burgio02} while the bag constant is different from ours. The
internal energy, which is the difference of the energy density with
respect to $\rho_B c^2$, of the mixed phase is larger than that of the
hadronic phase for high temperature regime ($T \ge 50$~MeV). This is not
unphysical because we should compare not the internal energies but the
free energies. In the left panels of FIG.~\ref{fg}, we show the
Helmholtz free energies per baryon as a function of the specific
volume. If the EOS is thermodynamically stable, this function is convex
downward. This feature is fulfilled for our models. In the right panels
of FIG.~\ref{fg}, on the other hand, we show the Gibbs free energies per
baryon as a function of the pressure. We can recognize that the free
energies of the mixed phase are always lower than those of the pure
hadronic and quark phases.

\begin{figure}
\begin{center}
\includegraphics{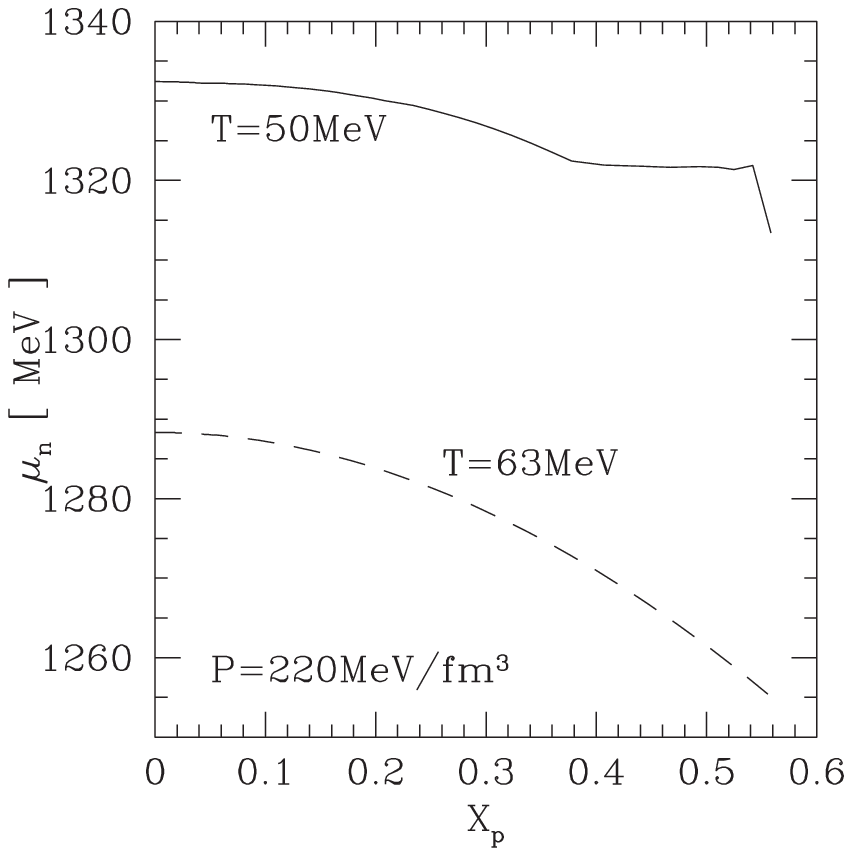}
\caption{Profiles of neutron chemical potential, $\mu_n$, as a
 function of ``total'' proton fraction, $X_p$, for fixed pressure
 $P=220$~MeV~fm$^{-3}$ and temperatures $T=50$~MeV (solid line) and
 63~MeV (dashed line). This is a result for the reference model of SN
 matter with $Y_l=0.1$.}
\label{prigof}
\end{center}
\end{figure}

For the phase transition of matter with two components, it is known that
the mixed phase should satisfy not only the condition of mechanical
stability, as verified above, but also the stability with respect
to diffusion \cite{prigo54}. In our model, the condition is expressed as
\begin{equation}
\left(\frac{\partial \mu_n}{\partial X_p}\right)_{P,T}\leq0,
\label{prigo}
\end{equation}
where $X_p$ is a ``total'' proton fraction and defined as
\begin{equation}
X_p=\frac{1}{n_B}\left(n_p+\frac{2n_u-n_d-n_s}{3}\right).
\label{prigo}
\end{equation}
This feature is shown for our results in FIG.~\ref{prigof} and satisfied
also in other ranges of the parameters. The reference model of the SN
matter with $Y_l=0.1$ for fixed pressure and temperature are indicated
in this figure. From these analyses, we can confirm that our EOS is
thermodynamically stable.

\begin{figure}
\begin{center}
\includegraphics{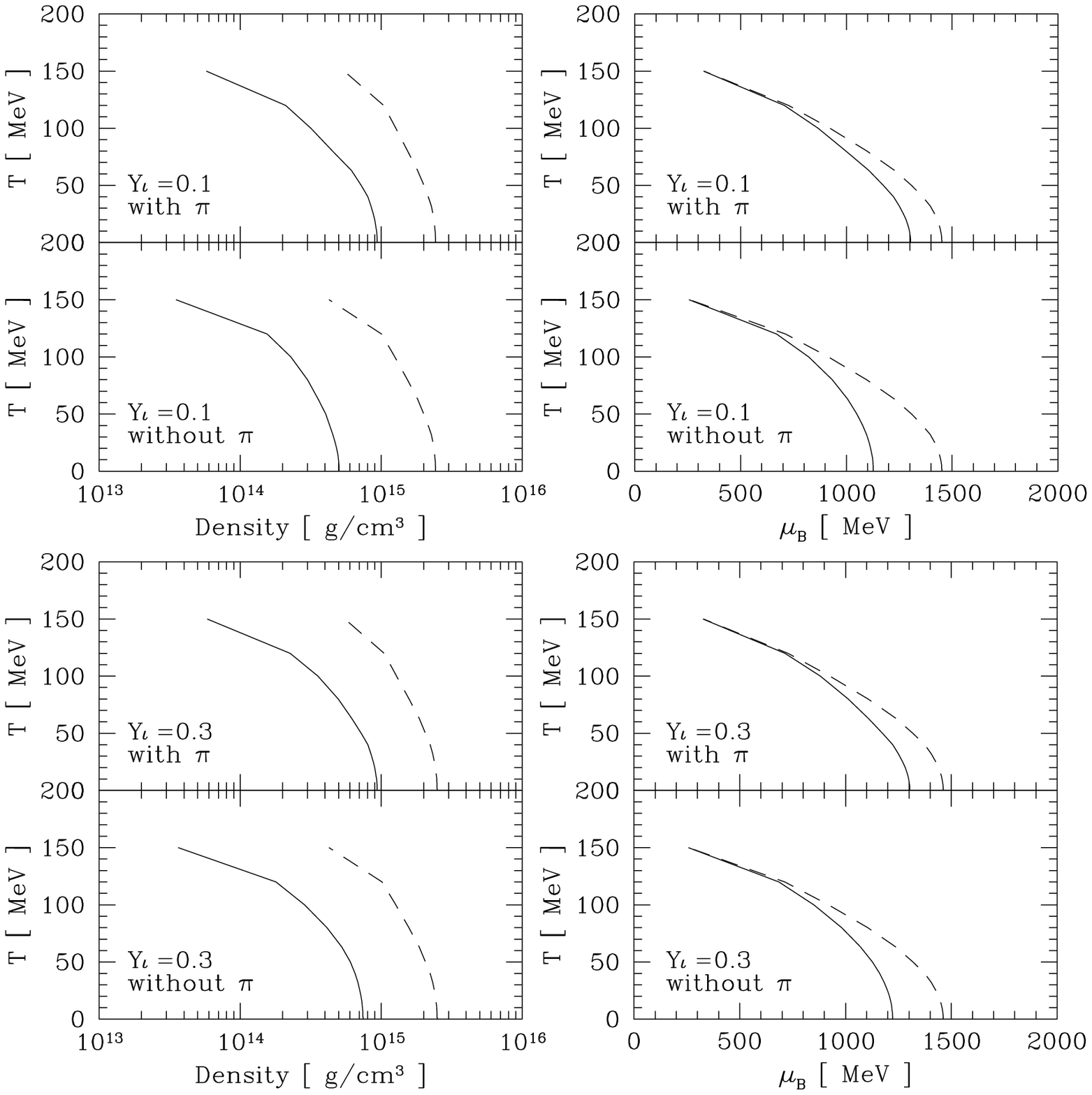}
\caption{Phase diagrams of our EOS of the SN matter with $Y_l=0.1$
 and 0.3 for $T<150$~MeV. Solid lines represent boundaries of hadronic
 matter and  mixed matter and dashed lines do those of mixed matter and
 quark matter. For the upper plots in each panel, an EOS with pions is
 used while an EOS without pions is used for the lower plots in each
 panel. $B=250$~MeV~fm$^{-3}$ is chosen for the bag constant for all
 panels.}
\label{temp}
\end{center}
\end{figure}

In FIG.~\ref{temp}, we show the temperature dependences of the
transition density and the critical baryon chemical potential for our
EOS of the SN matter with $Y_l=0.1$ and 0.3. These phase diagrams are
given not only for the reference model but also for a model without
pions while the bag constant of both models is $250$~MeV~fm$^{-3}$. The
baryon chemical potential in these figures, $\mu_B$, is the same as the
neutron chemical potential in the preceding sections, $\mu_n$. We can
see that the reference model has larger transition density and baryon
chemical potential than those of the model without pions. This is
because pions make the EOS softer; in other words, pressure gets lower
for a fixed baryon number density. Thus the condition of equilibrium
(\ref{atsuryoku}) is satisfied for larger density. Incidentally, the
phase diagram has been studied in Ref.~\onlinecite{drago99} while its
EOS for hadronic matter does not include pions. The color dielectric
model is used for the quark matter model whereas we use the MIT bag
model. In Ref.~\onlinecite{drago99}, the transition density gets lower
at higher temperature, which is the same as our result, however, its
transition density is lower than ours, especially at finite
temperature. For instance, the transition density is
$\sim10^{14}$~g~cm$^{-3}$ at $T\sim30$~MeV.

\begin{figure}[t]
\begin{center}
\includegraphics{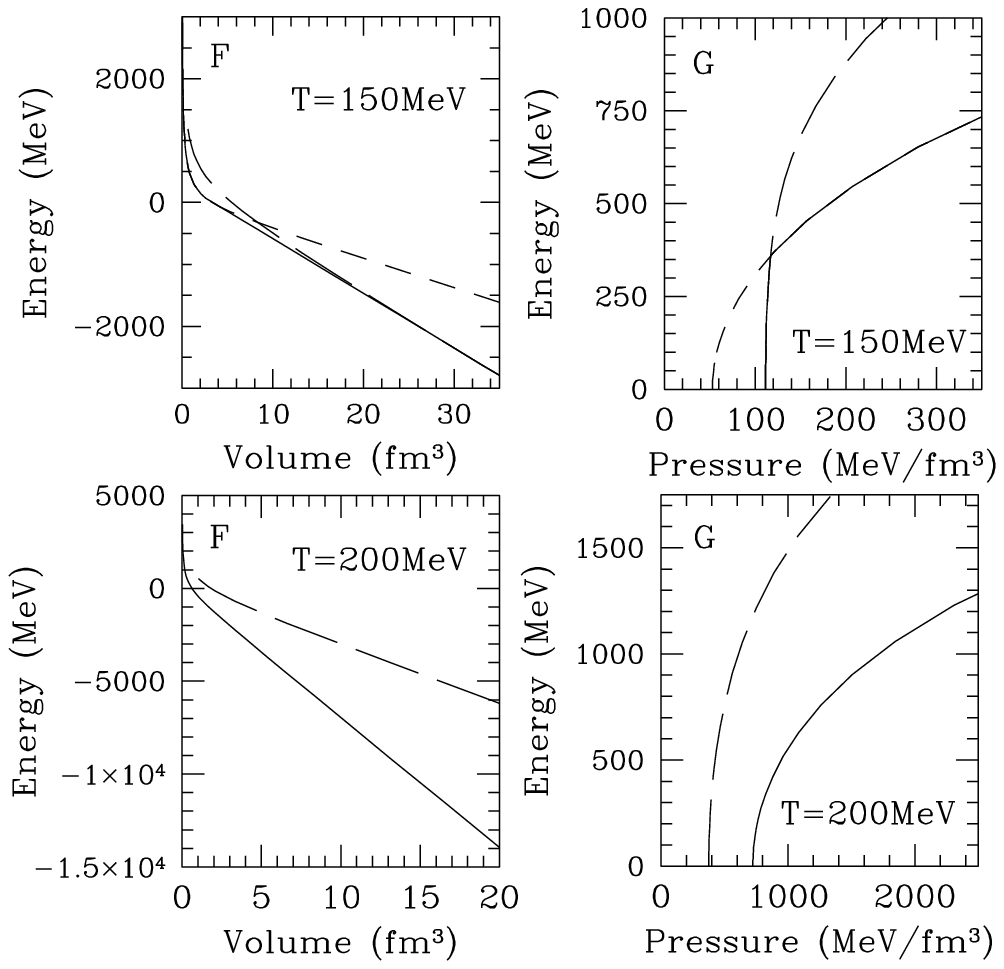}
\caption{Same as FIG.~\ref{fg} but for SN matter with $T=150$~MeV and
 $Y_l=0.1$ (upper panels), and for SN matter with $T=200$~MeV and
 $Y_l=0.1$ (lower panels).}
\label{fgh}
\end{center}
\end{figure}

It is suggested that the transition line has an end point in the high
temperature regime. The nature of the transition changes at this point,
while the exact phase diagram is not well known. This is a so-called
critical point. The temperature of the critical point, $T_c$, is
investigated experimentally by heavy-ion collisions \cite{star05,
phenix05} and theoretically by lattice QCD calculations \cite{bernard05,
cheng06, aoki06a, aoki06b}. From these studies, it may be in the range,
150~MeV~$\leq T_c \leq200$~MeV. Above the critical temperature, the
quark phase may be the most stable state for any densities. It is also
suggested that the baryon chemical potential of the critical point is
much smaller than the typical hadronic scale, $\mu_B \lesssim
40$~MeV. Although our model cannot describe the critical point in
detail, the critical baryon chemical potential drops dramatically with
the temperature in the regime $T\gtrsim100$~MeV. For much higher
temperature regime, the quark matter is more stable than the hadronic
matter even at zero density. As shown in FIG.~\ref{fgh}, for instance,
the hadron-quark phase transition occurs at $T=150$~MeV and the quark
phase is always the most stable state at $T=200$~MeV for the model with
$B=250$~MeV~fm$^{-3}$. Therefore, the plots for $T\gtrsim150$~MeV are
not given in FIG.~\ref{temp}. In this study, the bag constant $B$ is
independent of the temperature and the temperature at which the quark
matter becomes most stable depends solely on $B$. The larger the value
of $B$, the higher this temperature becomes. When this temperature is in
the range 150~MeV~$\leq T \leq200$~MeV, the bag constant is in
210~MeV~fm$^{-3}\lesssim B \lesssim 650$~MeV fm$^{-3}$
(200~MeV~$\lesssim B^{1/4} \lesssim 265$~MeV) for our models. It is
noted that the bag constant of the reference model,
$B=250$~MeV~fm$^{-3}$ ($B^{1/4}=209$~MeV), lies within this range.

\subsection{Maximum mass of the hybrid stars} \label{star}
\begin{figure}
\begin{center}
\includegraphics{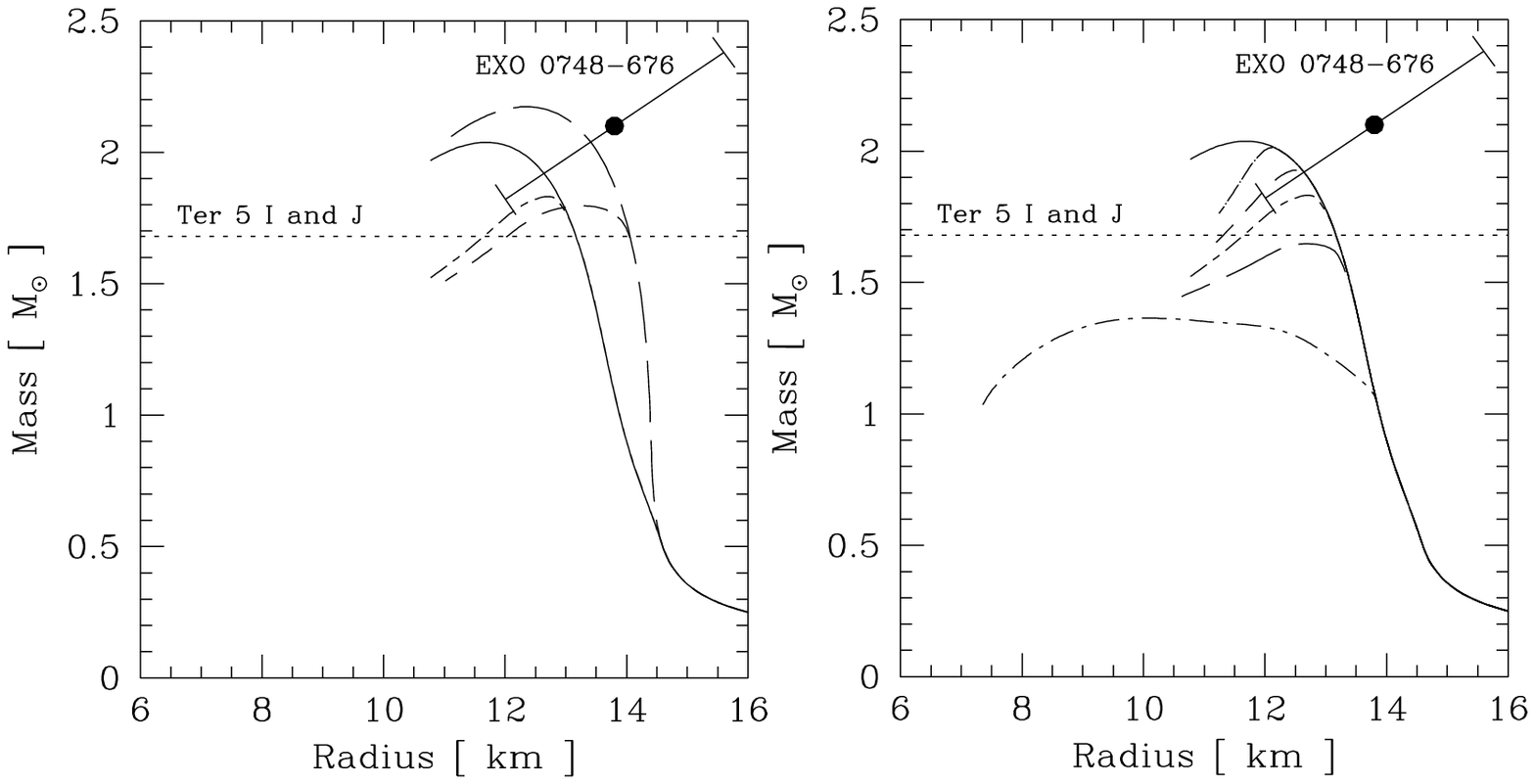}
\caption{Mass-radius trajectories for our EOS's of NS matter. In the
 left panel, long-dashed, short-dashed, solid and dot-dashed lines
 represent the models~OO (the original Shen EOS \cite{shen98a,
 shen98b}), OQ, PO and PQ (the reference model), respectively, where the
 definition of the models is given in the text. In the right panel, we
 show the results of the models with pions, and the solid line is the same
 as that in the left panel (PO). Other lines correspond, from bottom to
 top, to the models with pions and quarks with $B=150$, 200, 250, 300,
 400~MeV~fm$^{-3}$. In both panels, horizontal dotted lines represent
 the lower limit of the maximum mass of compact stars determined by
 pulsars Ter 5 I and J with 95\% confidence, and the plots with
 $1\sigma$ error bars are for the measurements of neutron star EXO
 0748-676.}
\label{maxmass}
\end{center}
\end{figure}

We examine the maximum mass of hybrid stars constructed by our EOS for
NS matter. In the following, we denote the EOS's without pions and
quarks (the original Shen EOS \cite{shen98a, shen98b}), without pions
but with quarks, with pions and without quarks and the model with pions
and quarks (the reference model) as OO, OQ, PO and PQ, respectively. The
bag constant is $B=250$~MeV~fm$^{-3}$ for the models with quarks. In
FIG.~\ref{maxmass}, we show the mass-radius trajectories for our EOS's
together with the recent data of compact star masses. We can see that
the pion population and the hadron-quark transition lower the maximum
mass because they soften the EOS. For instance, the maximum mass of
model~OO is 2.2 solar masses ($M_\odot$) while those of models~PO, OQ
and PQ are $2.0M_\odot$, $1.8M_\odot$ and $1.8M_\odot$, respectively. In
the right panel, we show the dependence on $B$. We can see that the
maximum mass becomes lower when the bag constant becomes small. This is
because the phase transition density is lower for smaller bag constants. 

The recent measurements of compact star masses increase the lower limit
of the maximum mass. For instance, it has been established with 95\%
confidence that at least one of the two pulsars, Ter 5 I and J, is more
massive than $1.68M_\odot$ from analysis of the joint probabilities
\cite{ranson05}. For the neutron star EXO 0748-676, it has been
reported that the lower limits on the mass and radius are
$M\geq2.10\pm0.28M_\odot$ and $R\geq13.8\pm1.8$~km with $1\sigma$ error
bars, while there are uncertainties in the analysis of the X-ray burst
spectra \cite{ozel06}. We note that the reference model
($B=250$~MeV~fm$^{-3}$) is consistent with these measurements whereas
models with lower bag constants $B\lesssim200$~MeV~fm$^{-3}$ in our
study produce lower maximum masses.

\subsection{Application to stellar core collapse} \label{collapse}
Recently, numerical studies on black hole formation by stellar core
collapse have been done for massive stars with an initial mass of
$\sim40M_{\odot}$ \cite{sumi06, sumi07, seki05, seki07}. The realistic
EOS and the realistic progenitor models are used in the studies of
spherically symmetric collapse in Refs.~\onlinecite{sumi06, sumi07}
whereas the parametric EOS and polytropic initial model are employed in
those of axisymmetric collapse in Refs.~\onlinecite{seki05,
seki07}. Collapse of very massive $\sim300M_{\odot}$ first generation
stars in the universe (so-called Population III stars) have been also
studied \cite{fryer01, self06, suwa07, liu07}. It is noted that these
studies do not include the hadron-quark phase transition. In this study,
we perform numerical simulations for a spherically symmetric stellar
collapse using the EOS's examined in the preceding section. We set the
bag constant $B=250$~MeV~fm$^{-3}$ for all the models with quarks. A
result of an evolutionary calculation for a Population III star with
$M=100M_{\odot}$ \cite{nomoto05} is chosen as the initial model of our
simulation. This model is the intermediate one between the models
stated above. The numerical simulation of its collapse has been already
done under the Shen EOS \cite{self07}.

We follow the scheme of Ref.~\onlinecite{self07} in our simulations. We
solve the general relativistic hydrodynamics and neutrino transfer
equations simultaneously under spherical symmetry \cite{yamada97,
yamada99}. Neutrino reactions are treated in detail
\cite{sumi05}. Incidentally, we have assumed for the EOS that the
electron-type neutrinos are in equilibrium with other particles in the
hadron-quark mixed phase and the pure quark phase. Hence after the phase
transition occurs, we do not compute the neutrino distribution functions
and assume that they are Fermi-Dirac functions for all species
conserving the electron-type lepton fraction, $Y_l$, for each fluid
element. Moreover, we neglect the entropy variation from the neutrino
transport. We can justify this gap in the neutrino treatments because
the density is high enough for neutrinos to be trapped anyway at the
phase transition. It is also noted that we can compute the collapse up
to the apparent horizon formation, which is a sufficient condition for
the formation of a black hole. For the details of our simulation such as
the initial condition, numerical methods and convergences, see
Ref.~\onlinecite{self07}.

\begin{figure}
\begin{center}
\includegraphics{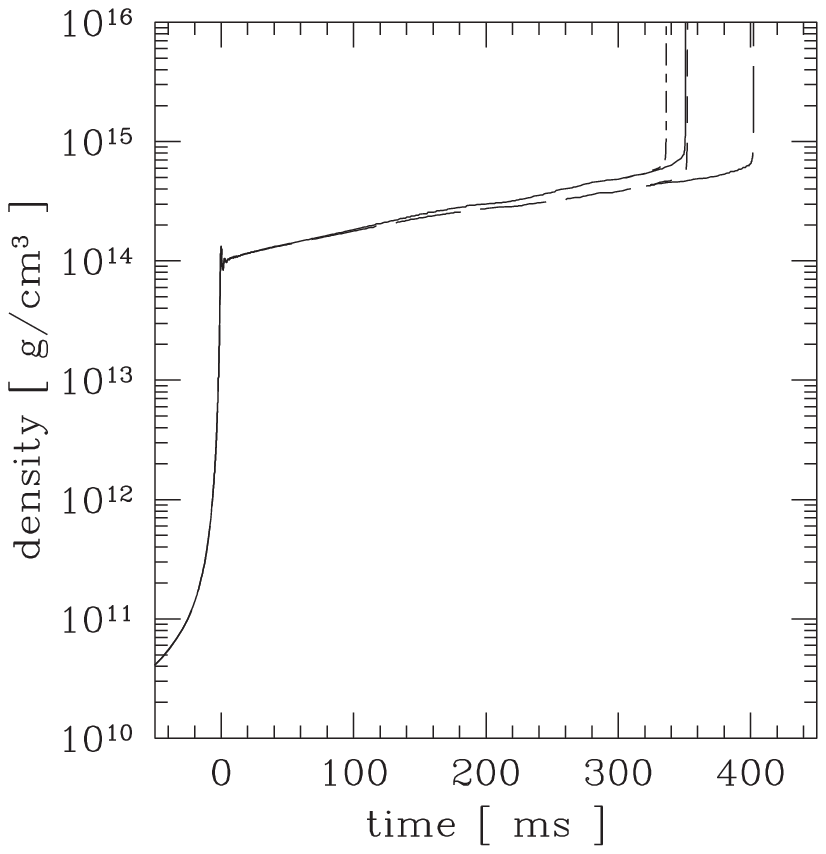}
\caption{Time profiles of the central baryon mass density. The notation
 of the lines is the same as in the left panel of FIG.~\ref{maxmass} and the
 time is measured from the point at bounce.}
\label{dens}
\end{center}
\end{figure}

In FIG.~\ref{dens}, we show the time profiles of the central baryon mass
density. These models have a bounce owing to thermal nucleons at 
subnuclear density ($\sim1.4\times10^{14}$ g cm$^{-3}$) and then
recollapse to a black hole. The contribution of pions makes a difference
at $\sim2\times10^{14}$~g~cm$^{-3}$ and that of quarks does so for
larger density. We can also see that the effect of quarks begins to work
suddenly at the transition density whereas that of pions does
gradually. This is because the thermal pions appear before the pion
condensation. We note that the duration of neutrino emission is almost
the same as the interval time from the bounce to the apparent horizon
formation \cite{sumi06, sumi07}. In Table~\ref{neuene}, we show them for
all models computed here. Since the EOS becomes softer owing to the
contribution of pions and quarks, the interval time becomes
shorter. Model~OO (the original Shen EOS) takes 20\% longer to
recollapse than model~PQ (the reference model in this study) does. Using
this difference of the interval time, we may be able to probe
observationally the EOS of hot dense matter in future.

\begin{table}[t]
\caption{Results of the numerical simulations. The definition of the
 models is given in the text (Sec.~\ref{star}). $t_\mathrm{rec}$
 represents the interval time from the bounce to the apparent horizon
 formation. $E_{\nu_i}$ is the total energy of emitted $\nu_i$, where
 $E_{\nu_x}=E_{\nu_\mu}=E_{\bar{\nu}_\mu}=E_{\nu_\tau}=E_{\bar{\nu}_\tau}$.
 $E_\mathrm{all}$ is the total energy summed over all species.}
\begin{center}
\setlength{\tabcolsep}{15pt}
\begin{tabular}{lccccc}
\hline\hline
 Model & $t_\mathrm{rec}$ (msec) & $E_{\nu_e}$ (ergs) & $E_{\bar \nu_e}$ (ergs) & $E_{\nu_x}$ (ergs) & $E_\mathrm{all}$ (ergs) \\ \hline
 OO & 402 & $9.42\times10^{52}$ & $7.89\times10^{52}$ & $4.40\times10^{52}$ & $34.9\times10^{52}$ \\
 OQ & 352 & $8.03\times10^{52}$ & $6.58\times10^{52}$ & $3.44\times10^{52}$ & $28.4\times10^{52}$ \\
 PO & 351 & $8.05\times10^{52}$ & $6.63\times10^{52}$ & $3.67\times10^{52}$ & $29.3\times10^{52}$ \\
 PQ & 336 & $7.60\times10^{52}$ & $6.22\times10^{52}$ & $3.34\times10^{52}$ & $27.2\times10^{52}$ \\
\hline\hline
\end{tabular}
\label{neuene}
\end{center}
\end{table}

The total neutrino energies emitted from each model are shown in
Table~\ref{neuene}. In this simulation, we assume that $\nu_\tau$
($\bar\nu_\tau$) is the same as $\nu_\mu$ ($\bar\nu_\mu$), and the
luminosities of $\nu_\mu$ and $\bar\nu_\mu$ are almost identical. This
is because they have the same kind of reactions and the difference of
coupling constants is minor. Therefore we denote these four species as
$\nu_x$ collectively in Table~\ref{neuene}, taking the
average. Comparing models OO and OQ or models PO and PQ in
Table~\ref{neuene}, we can see that the total energies of emitted 
neutrinos for the models including quarks become lower than those of
the models without quarks. This is because the collapse is hastened by
quarks, and their effects make differences only for the last moment
as discussed above. Thus the duration of the neutrino emission becomes
shorter while the neutrino luminosity does not change much.

This trend can be seen in the effects of pions also. However, the
neutrino luminosity for the models with pions differs gradually in time
from that for the models without pions because the effect of pions works
gradually. The neutrino luminosity summed over all species is given
approximately by the accretion luminosity  $L^\mathrm{acc}_\nu \sim G
M_\nu \dot{M} / R_\nu$ \cite{thomp03}, where $G$, $R_\nu$, $\dot{M}$ and
$M_\nu$ are the gravitational constant, the radius of the neutrino
sphere, the mass accretion rate and the mass enclosed by $R_\nu$,
respectively. We can roughly regard the neutrino sphere as the surface
on which neutrinos are emitted, and $R_\nu$ is defined as,
\begin{equation}
\int^{R_s}_{R_\nu} \frac{\mathrm{d}r}{l_\mathrm{mfp}(r)} = \frac{2}{3},
\label{nsphere}
\end{equation}
where $R_s$ is the stellar radius and $l_\mathrm{mfp}$ is the mean
free path of neutrino. Incidentally in our case, as is the case for an 
ordinary supernova, the proto-neutron star is formed before the black
hole formation, and neutrinos are emitted mainly on the surface of the
proto-neutron star. Thus we can regard $R_\nu$ and $M_\nu$ as the radius
and mass of the proto-neutron star, respectively. On the other hand, the
density of the proto-neutron star of the model with pions is greater than
that of the model without pions comparing at the same time because the
EOS becomes soft under the influence of pions as seen in
FIG.~\ref{dens}. Thus the radius of the proto-neutron star becomes
smaller without changing the mass of it. As a result, the neutrino
luminosity gets higher for the models including pions. For instance, at
300~ms after the bounce, the luminosity of model~OO is
$1.09\times10^{53}$~erg~s$^{-1}$ whereas that of model~PO is
$1.27\times10^{53}$~erg~s$^{-1}$. While a simple comparison cannot be
done because the quarks also affect the neutrino emission slightly, we
can see that model~PO has higher total energy than that of model~OQ
although the interval times are almost the same. In conclusion, pions
make the total energy of the emitted neutrinos lower by shortening the
interval time and higher by increasing luminosity.

The computations shown here are done for only a single model,
$M=100M_{\odot}$ and the bag constant $B=250$~MeV~fm$^{-3}$. For the
models with a smaller bag constant, an interval time of the neutrino
emission may become much shorter because of the smaller maximum mass of
the hybrid stars. It is noted that we are preparing more detailed
studies for various cases of stellar collapse \cite{self08a, self08b}.

\section{Summary} \label{summary}
We have constructed the EOS including the hadron-quark phase transition
using Gibbs conditions for finite temperature. As for the hadronic
phase, we have added pions to the EOS by Shen~et~al. \cite{shen98a,
shen98b}, which is based on the relativistic mean field theory. Our
results are found to reproduce the composition in the previous studies
particularly for the pion population in the neutrino-less $\beta$
equilibrium state at zero temperature \cite{glende85}. As for the quark
phase, we have adopted the MIT bag model of the deconfined 3-flavor
strange quark matter \cite{bag74} and the bag constant has set to be
$B=250$~MeV~fm$^{-3}$ for the reference model. As for the mixed phase,
we have assumed that the equilibrium is achieved not only by the strong
interactions but also by the weak interactions. Electrons and
electron-type neutrinos are taken into account as being in weak
equilibrium for the mixed phase and pure quark phase. Incidentally, we
have treated the muons approximately. Our EOS is thermodynamically
stable and exhibits qualitatively the desired properties of hadron-quark
mixed matter, such as the temperature dependence of the transition
density.

We have studied the astrophysical implications of the calculated
EOS. The most important effect of pions and quarks is to soften the
EOS. Pions and quarks make the maximum mass of compact stars smaller. It
is noted that the models including pions and quarks with the bag
constant $B\lesssim200$~MeV~fm$^{-3}$ give the maximum mass that is
smaller than the recent measurements of compact stars. We have computed
the collapse of a massive star with $100M_{\odot}$ using our EOS and
found that the interval time from the bounce to the black hole (apparent
horizon) formation becomes shorter by the inclusion of pions and
quarks. As a result, they affect the total energy of the emitted
neutrinos because of the shorter duration. We have also shown that the
neutrino luminosity becomes higher under the effect of pions because
they raise the density of the proto-neutron star core. We stress that we
may be able to discuss the EOS of hot dense matter detecting the
neutrinos from black hole progenitors in future.

The current paper is the first attempt to clarify the effects due to the
quark-hadron phase transition on the black hole formation and the
neutrino emission. We are planning to study the initial mass dependence
\cite{self08a} and the bag constant dependence \cite{self08b} of stellar
core collapse systematically. We will also investigate effects of
hyperons in the hadronic matter \cite{self08b}. There are more problems
to be solved regardless of our project. For instance, other quark models
like a color superconducting phase and effects of surface tension and
screening of charged particles on the equilibrium condition in the mixed
phase are worth investigating.

\begin{acknowledgments}
 We are grateful to Hideyuki Umeda for providing a progenitor model. We
 would like to thank Akira Ohnishi, Yoshiaki Koma and Chikako Ishizuka for
 fruitful discussions. Numerical computations were performed on the 
 Fujitsu VPP5000 at the Center for Computational Astrophysics (CfCA) of
 the National Astronomical Observatory of Japan (VPP5000 System projects
 ikn18b, ukn08b, iks13a, uks06a), and partially on the supercomputers in
 JAERI, YITP and KEK (KEK Supercomputer project 108). This work was
 partially supported by Japan Society for Promotion of Science (JSPS)
 Research Fellowship, Grants-in-Aid for the Scientific Research from the
 Ministry of Education, Science and Culture of Japan (14740166,
 15540243, 15740160, 17540267, 18540291, 18540295, 19540252) and the
 21st-Century COE Program ``Holistic Research and Education Center for
 Physics of Self-organization Systems.''
\end{acknowledgments}

\bibliographystyle{apsrev} 
\bibliography{apssamp}

\end{document}